# On non-linear excitation of voids in dusty plasmas

**E.Nebbat[1], R. Annou[2,3], and R. Bharuthram[2]**

[1]Faculty of Physics, USTHB. Algiers (Algeria).

[2] School of Pure and Applied Physics, UKZN. Durban (South Africa).

[3] Permanent address: Faculty of Physics, USTHB. Algiers (Algeria).

**Abstract**

The void, which is a dust-free region inside the dust cloud in the plasma, results from a balance of the electrostatic force and the ion drag force on a dust particulate. The ion drag force having numerous forms, some of which are based on models whereas others are driven from first principles. To explain the generation of voids, Avinash *et al.* proposed a time-dependent nonlinear model that describes the void as a result of an instability. We augment this model by incorporating the grain drift and by replacing the ion-drag force by that derived by Khrapak *et al.*, in a spherical configuration. It has been revealed that the void formation is a threshold phenomenon, i.e., it depends on the grain size. Furthermore, the void possesses a sharp boundary beyond which the dust density goes down and may present a corrugated aspect. For big size grains, the use of both ion drag forces leads to voids of a same dimension, though for grains of small sizes the Avinash force drives voids of a higher dimension.

## §.I. Introduction

Dusty plasmas are plasmas containing micron or submicron dust grains that have very high charge and mass. These plasmas may be encountered in a wide range of situations spanning from astrophysical to industrial situations **[1].** The charged dust grains exhibit necessarily a collective behavior due to their charges **[2]**. Indeed new modes have been predicted and proved to exist by way of numerous experiments **[3]**. In addition to the dust grain dynamics, the charge fluctuations have been shown to affect the existing modes supported by the plasma. Recently, the grain mass fluctuations caused by the grain sputtering by the plasma ions have been proved to influence the dielectric properties of the plasma as well **[4]**. Generally, this influence of dust grains on the plasma may be considered spurious in many cases such as the semi-conductor manufacturing, as it leads to material yield loss. A method has been proposed to push away dust grains from the center of the plasma, by using a ponderomotive force generated by an appropriate wave **[5]**. In this case, the grain size should be small enough as a big grain would have a gyro-radius greater than the experimental set-up dimension. However, in many applications dust particles may be highly desirable such as special material processing, e.g, nanocristalline powders of pure metals **[6]**. In addition to the modification of the wave spectra, the grains in a plasma exhibit a spectacular phenomenon that has been discovered the last decade, namely, the dust void. As a matter of fact, it has been reported by Praburam and Goree **[7]**, that when particulates grow to a critical size, a cloud is formed in the interelectrode zone. Later on, the cloud undergoes a transition to a turbulent state, where a very low-frequency filamentary mode is excited. After this mode developed, another mode coined by the authors "great void" appeared. It is a rotating region free from dust grains. It has been stated that the shape of the void is determined by the combined effect of the electric force and the ion drag force that expresses the momentum transfer from collisions of flowing ions with the grain. Later, Samsonov and Goree **[8]** conducted more sophisticated experiments to support the previous conclusion. Indeed, according to the authors, when the grain reaches the critical value the ion drag force overcomes the electric force and the grains are moved outwardly. Hence, the electrons undergo less depletion, so ionization is enhanced (and so is the glow). It appears then, that the void is characterized by a critical size of the dust grain and a sharp boundary. It should be noted that in the same time, experiments on voids have been conducted also by Melzer *et al.* **[9]**. Nevertheless, the lack of a non-linear theory that explains the critical size of the grain, the sharp boundary of the void as well as its size remained. To this end, Goree *et al.* **[10]** developed a one-dimensional non-linear fluid model to explain the formation of the void and its properties, where dust charging, ion drag force on dust, and a source of ionization have been taken into account. It has been found that a void represents a stable equilibrium in a plasma with a sharp boundary where an ionization source is a necessary condition. Unfortunately, and according to Avinash *et al.***[11]**, though the proposed interpretation of the void formation has been confirmed by numerous experiments, non-linear time-dependent models that cope with the spontaneous development of the linear instability and the appearance of the void in the final stage, are still lacking. Hence they proposed a time-dependent self-consistent non-linear model for void formation, which is a one-dimensional model that incorporates grain diffusion, collisions with neutrals, and an approximate expression of the non-linear ion drag force (ABH force). In this paper, we augment the work of Avinash *et al.* by including three elements, viz., i/ the grain drift, ii/ a more accurate ion drag force, we use in particular the expression derived by Khrapak *et al.* ( KIMT force) **[12]**, and iii/ we consider the void formation in a spherical configuration (dependence on r), in which case, the three dimensional picture may be achieved by allowing rotational symmetry that is not allowed for a one-dimensional solution (dependence on $x$). The paper is organized as follows, in Sec. I we introduce the subject; the formulation of the problem being exposed in Sec. II, whereas in Sec. III the results are discussed. Sec. IV is devoted to the concluding remarks.

## §.II. Formulation

Let us consider an electron-ion plasma containing a cloud of dust grains. The evolution of this system is described by the continuity and momentum equations, which are coupled to Poisson's equation. The dust continuity equation is written as,

$$\frac{\partial n}{\partial t} + \nabla.(n\vec{\mathrm{v}}) = D\,\nabla^2 n - \frac{Z\,e\,D}{T}\nabla\left(n\vec{E}\right), \qquad \textbf{(1)}$$

where the second term on the RHS of Eq. (1), corresponds to the grain drift under the action of the electric field $\vec{E}$, and $n$, $\vec{\mathrm{v}}$, $T$, $Z$, and $D$, are respectively the dust density, dust velocity, dust temperature, dust charge and dust diffusion coefficient.
The dust momentum equation is given by,

$$m\left[\frac{\partial\vec{\mathrm{v}}}{\partial t} + \left(\vec{\mathrm{v}}.\vec{\nabla}\right)\vec{\mathrm{v}}\right] = -Ze\vec{E} - m\,v_{dn}\,\vec{\mathrm{v}} - \frac{\vec{\nabla}P}{n} + \vec{F}_d \qquad \textbf{(2)}$$

where, $P$ is dust pressure and $v_{dn}$ is the dust-neutral collision frequency. We deal with two expressions of the non-linear ion drag force, namely, i/ the ABH form that is given by, $F_d = m\, v_{di}\, \mathrm{v}_{\mathrm{thi}} \dfrac{u}{b+u^3}$, where $v_{di}$ is the dust-ion collision frequency, $\mathrm{v}_{thi}$ is the ion thermal velocity, $b$ is a fitting constant, and $u = \dfrac{\mathrm{v}_i}{\mathrm{v}_{thi}}$, $\mathrm{v}_i$ being the ion velocity that is given by, $\mathrm{v}_i = \dfrac{e\,E}{m_i\, \nu_{in}}$, where $v_{in}$ is the ion-neutral collision frequency, and ii/ the KIMT form, which is cast as, $F_d = \dfrac{8\sqrt{2\pi}}{3} r_d^{\,2}\, n_i\, m_i\, \mathrm{v}_{thi}\, \mathrm{v}_{\mathrm{i}} \left(1 + \dfrac{1}{2} z\, \tau + \dfrac{1}{4} z^2\, \tau^2\, \Lambda\right)$, where, $r_d$ is the grain radius, $z = \dfrac{Z\, e^2}{r_d\, T_e}$, $\tau = \dfrac{T_e}{T_i} = \dfrac{1}{\tau_i}$ and the modified Coulomb logarithm is given by $\Lambda \approx 2\, F\left[\dfrac{\beta(\mathrm{v}_{thi})}{2}\right]$, where $F(x) = \int_x^{\infty} \dfrac{\exp(x-t)}{t}\, dt$, and $\beta(\mathrm{v}_{thi}) \approx \dfrac{z\, \tau\, r_d}{\lambda_D}$, $\lambda_D$ being the plasma Debye length.

As far as the electrons are concerned the electron momentum equation is cast as,

$$\vec{0} = -\,e\vec{E} - \frac{\vec{\nabla} P_e}{n_e}, \qquad \mathbf{(3)}$$

where $P_e$ and $n_e$ are the electron pressure and density.

The Poisson's equation is given by,

$$\nabla \vec{E} = 4\pi\, e\left(n_i - n_e - Z\, n\right) \qquad \mathbf{(4)}$$

Equations (1-4) are satisfied identically for a homogeneous equilibrium with $E = 0$, $\mathrm{v} = 0$ and $n_0 = \dfrac{n_{i0} - n_{e0}}{Z}$. By performing a stability analysis, that is perturbing the system around the equilibrium state and assuming an $\exp(\omega t + kr)$ dependence for the perturbed quantity, we may find the eigen values (ABH form) in the case $D = k = 0$ , as follows,

$$\omega_{1,2} = \frac{1}{2}\left\{ -\nu_{dn} \pm \sqrt{\nu_{dn}^2 + 16\pi\, n\, Z^2\, e^2 \left( -1 + \frac{a}{b} \right)} \right\}, \qquad \textbf{(5)}$$

where, $a = \dfrac{m\, v_{di}}{m_i\, Z\, v_{in}}$. It is clear that the equilibrium corresponding to ABH form of the ion drag force is unstable to purely growing modes when $\frac{a}{b} \rangle 1$. Hence the void formation is a threshold phenomenon. By introducing the grain radius in the inequality through the quantities, $m = \frac{4}{3}\pi\ \rho\, r_d^{\ 3}$ , where $\rho$ is the grain mass density, $Z = \dfrac{4\, T_e}{e^2}\, r_d$ for an argon plasma and $v_{di} = n_i\, \pi \sqrt{\dfrac{3\, T_i}{m_i}}\, r_d^{\ 2}$, one finds that the onset of the instability, or in other words, the void is formed when the grain radius exceeds the value,

$$r_d^{\ *} = \frac{3\, T_e\, b\, m_i\, v_{in}}{\pi^2\ \rho\, n_i\, e^2} \sqrt{\frac{m_i}{3\, T_i}} \qquad \textbf{(6)}$$

In Fig(1), we show that the onset of the instability is assured for $r_d \rangle r_d^{\ *}$ . In an argon plasma defined quantitatively by the following set of values, i.e, $m_i = 6.623\, 10^{-23}\ g$ ; $b = 1.6$ ; $\nu_{in} = 10^6\ s^{-1}$ ; $T_i = 0.015\, eV$ ; $n_i = 1.86\, 10^9\ cm^{-3}$ ; $n_0 = 2.37\, 10^4 cm^{-3}$, $T_e = 3\ eV$ and $\rho = 2\ g\ cm^{-3}$, we found $r_d^{\ *} \approx 58\, nm$ in the case of ABH form of the ion drag force. In Figures (2) and (3) we plotted the numerical density in the saturated state of the instability versus the position. It is clear that for $r_d \approx 50\ nm$, there is no void with a sharp boundary whereas for $r_d \approx 120\ nm$, the void with a sharp boundary appears.

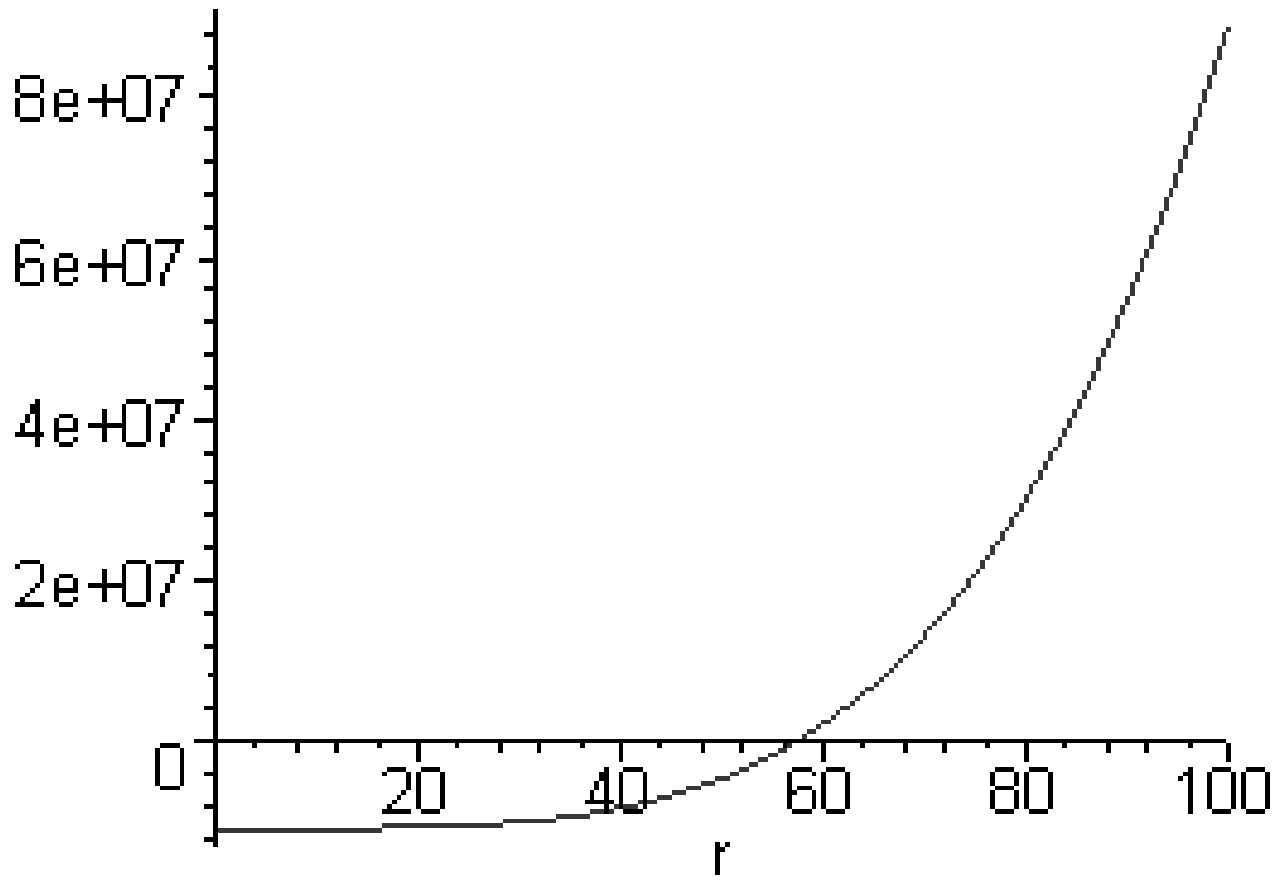

Fig. 1. $\left(\frac{a}{b}-1\right)$ versus the grain radius $r_d$

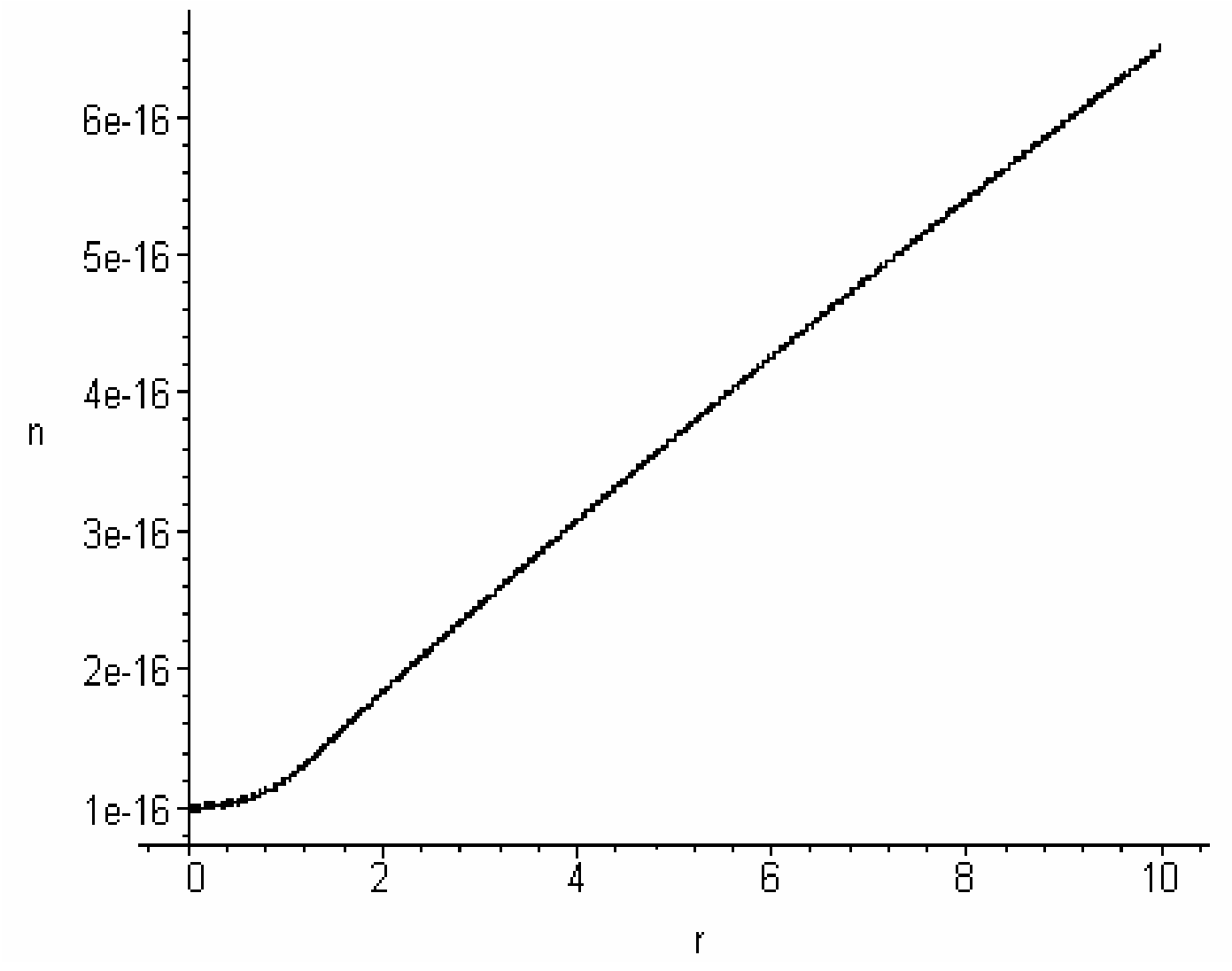

Fig. 2. The dust numerical density $n$ versus the position $r$ for $r_d = 50\ nm$

(ABH ion drag force)

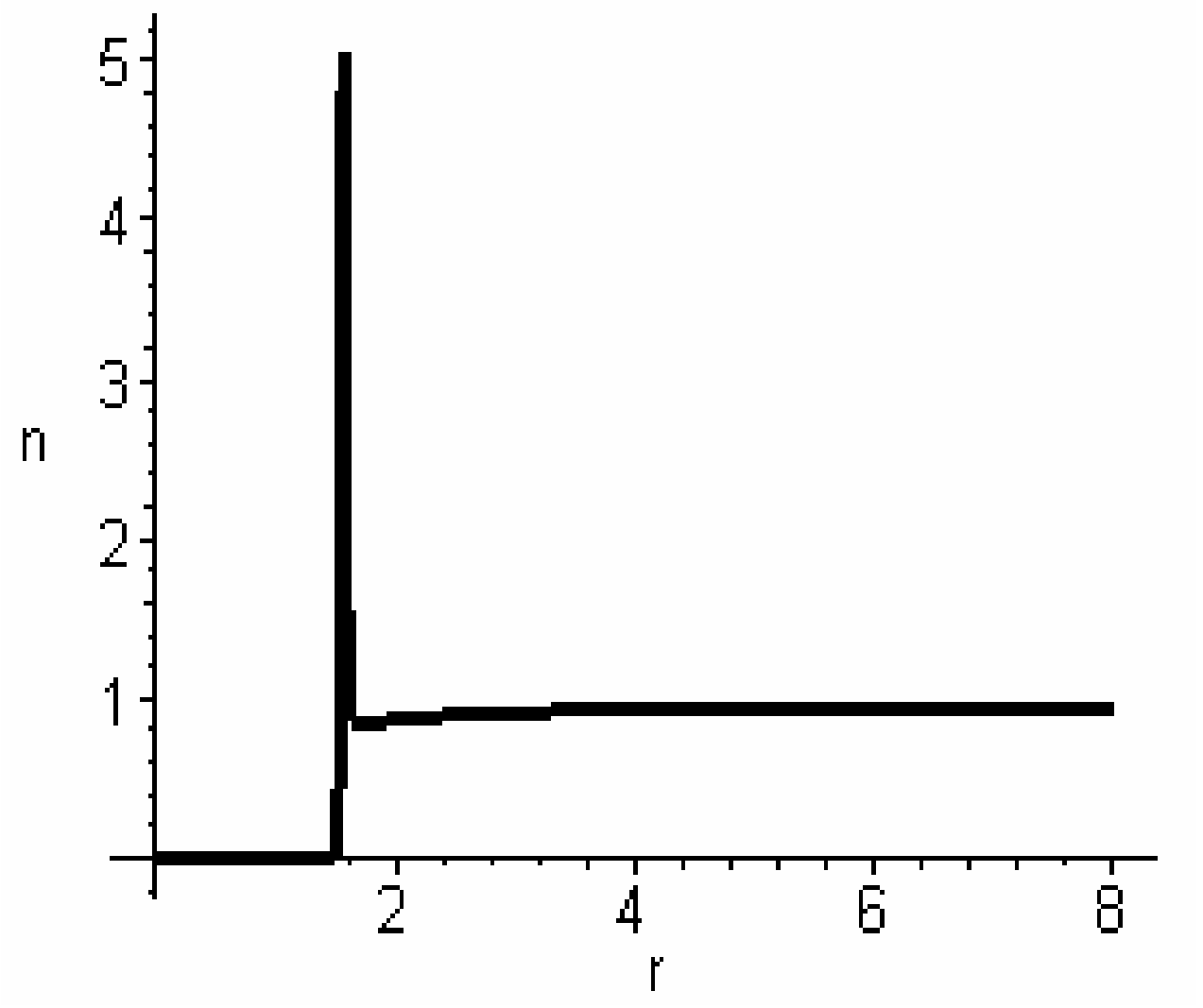

Fig .3. The dust numerical density $n$ versus the position $r$ for $r_d = 120\,nm$

(ABH ion drag force)

On the other hand, when the KIMT form of the ion drag force is considered, the condition for the onset of the instability is given by,

$$f(r_d) = \left(1 + \frac{2\,T_e}{T_i}\right) r_d + \left(\frac{2\,T_e}{T_i}\right)^2 r_d\, F\left(\frac{2\,T_e}{T_i}\, r_d\right) - \frac{3}{2}\,\frac{v_{in}\, T_e}{\sqrt{2p}\; n_i\, \mathrm{v}_{\mathrm{thi}}\, e^2} \rangle\, 0 \qquad \textbf{(7)}$$

We plotted this function in Fig.4, where it appears that the critical size of the grain is given by $r_d^{\,*} = 280\,nm$, which is clearly a more stringent condition to be fulfilled by the grain, as compared to the case of ABH form of the ion drag force.

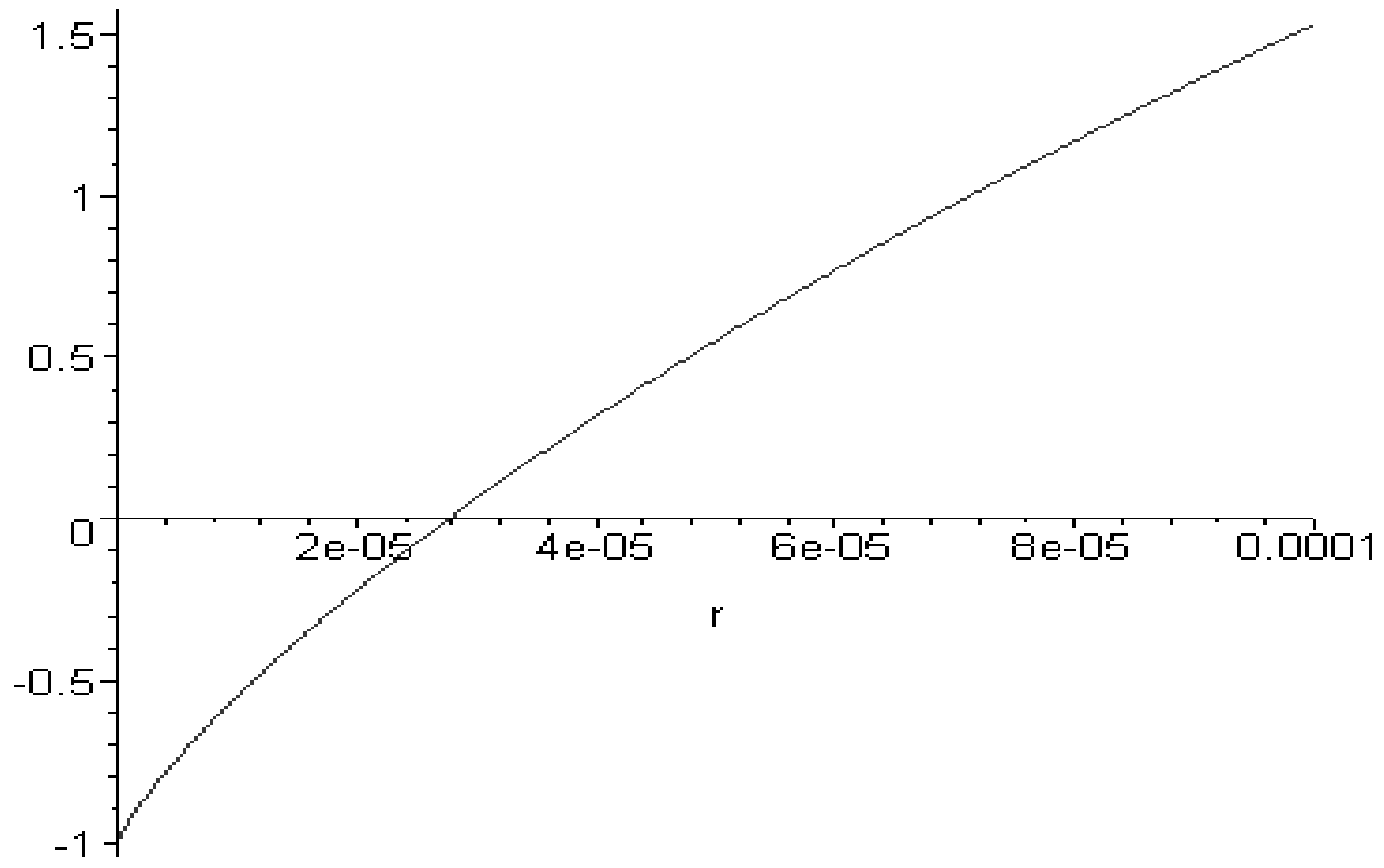

Fig.4. $f(r_d)$ versus the grain radius $r_d$

Figures (5) and (6) are two examples corresponding to two radii of the grain, one below the critical value ($r_d = 250\ nm$) and the other above this value ($r_d = 350\ nm$), where the void is clearly shown. This void is 100 % free of dust particles and possesses a sharp boundary as already described by numerous experiments.

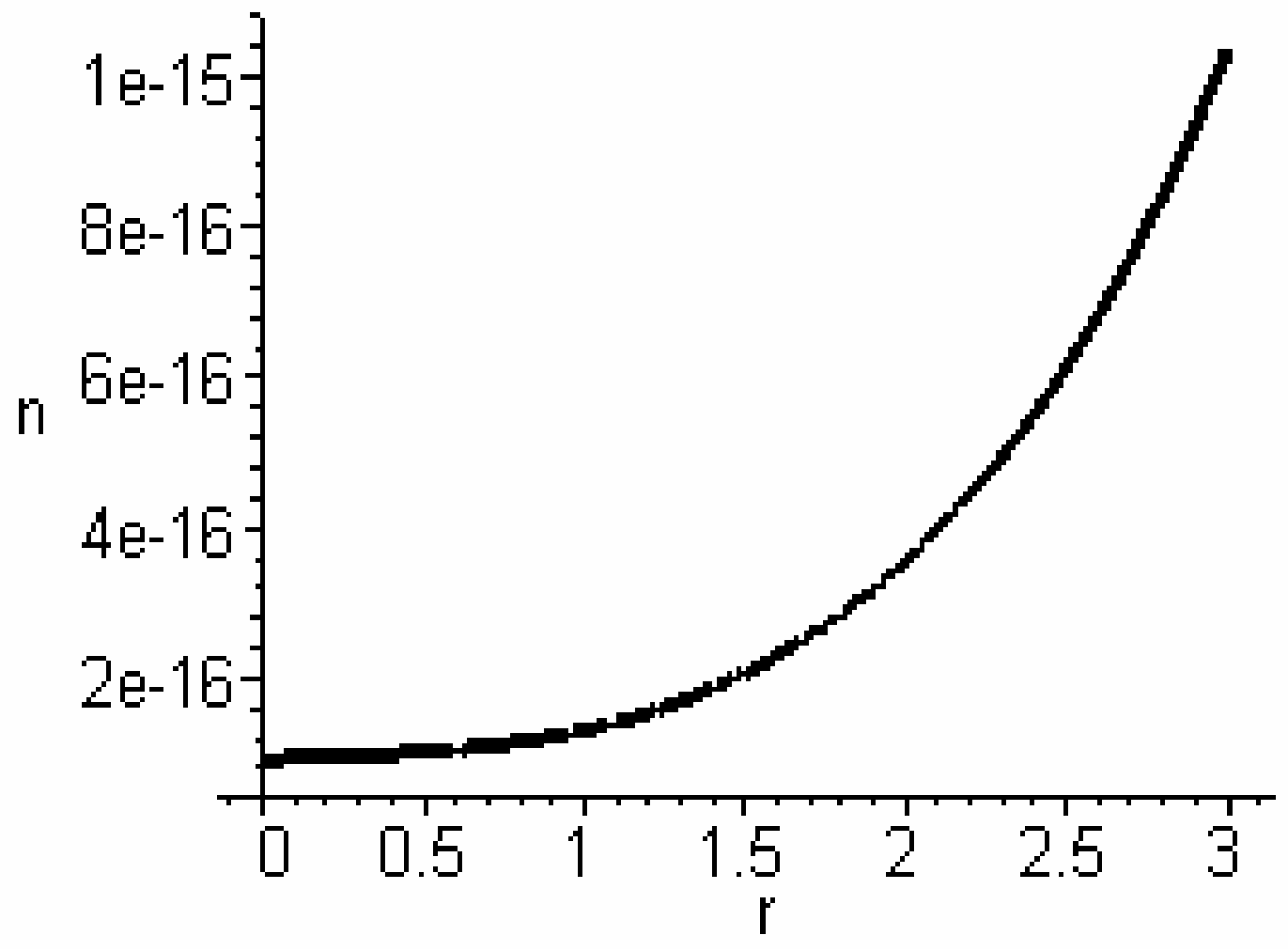

Fig.5. The numerical dust density $n$ versus the position $r$ for $r_d = 250\ nm$ (KIMT ion drag force)

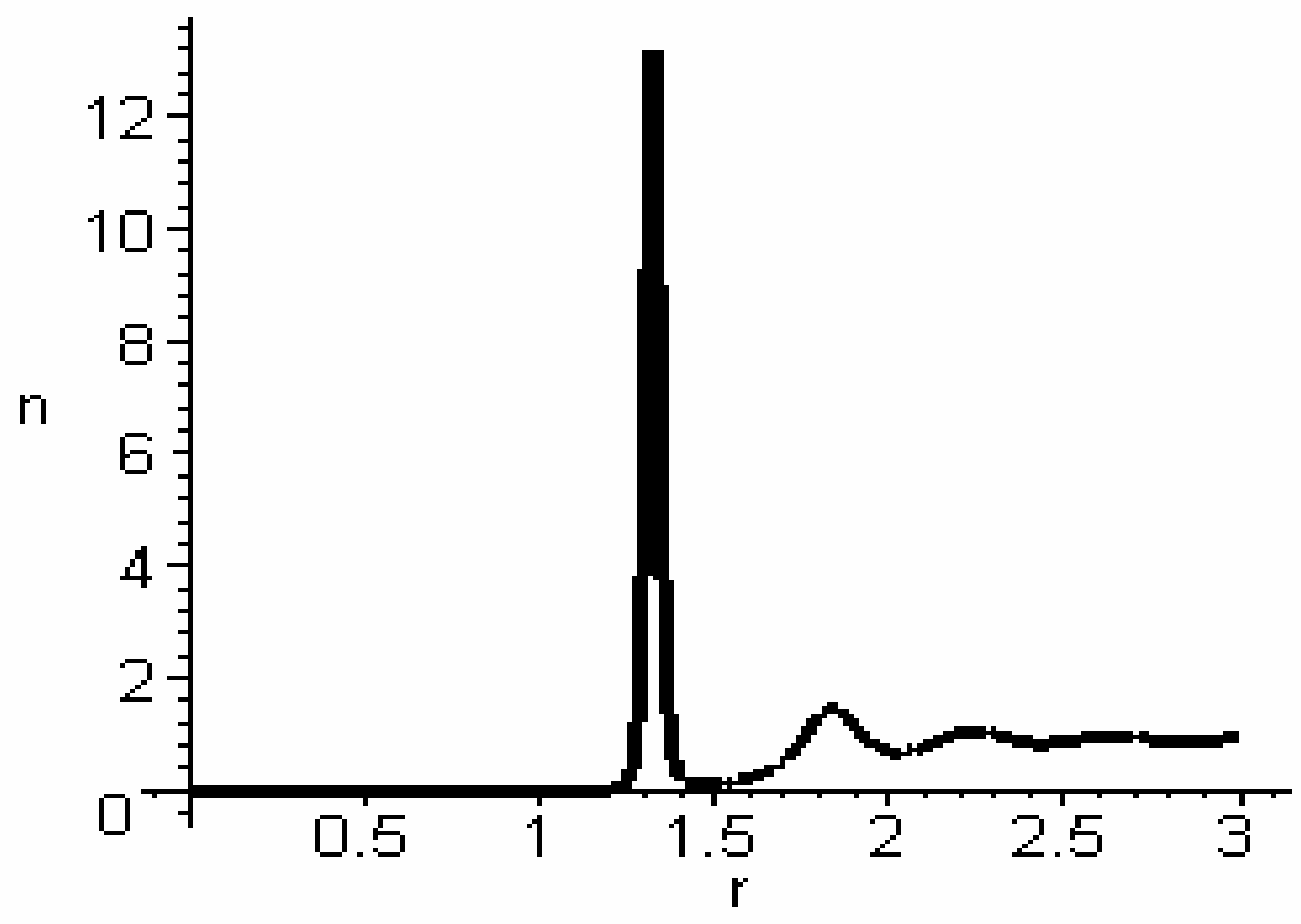

Fig.6. The numerical dust density $n$ versus the position $r$ for $r_d = 350\ nm$ (KIMT ion drag force)

In order to investigate the non-linear system of equations (1-4) governing the plasma evolution, we rewrite it in a dimensionless form as follows,

$$\frac{\partial n}{\partial r} = \Psi \, , \tag{8}$$

$$\frac{\partial \Psi}{\partial r} = \frac{n}{D\,\mathrm{v}}\left\{ c\left(\frac{F_d}{Z\,e} - E\right) + \mathrm{v}\,h\left(-n_e - n + 1\right) - \mathrm{v}\,a + \Psi\left[-\frac{\mathrm{c}}{\mathrm{Z\,n}} + \frac{\mathrm{v}}{\mathrm{n}}\left(\mathrm{v} + h\,E\right) - \frac{2\,t_i\,D}{l_{di}\,r}\right] + \frac{2\,t_d\,t_i\,\mathrm{v}^2}{\mathrm{Z\,r}}\right\} \, , \tag{9-a}$$

$$\frac{\partial \mathrm{v}}{\partial r} = \frac{c}{\mathrm{v}}\left(-E + \frac{F}{Z\,e}\right) - \frac{c\,m\,t_d\,t_i\,\Psi}{n\,\mathrm{v}} - a \, , \tag{10-a}$$

$$\frac{\partial n_e}{\partial r} = -\frac{n_e\,E}{t_i} \, , \tag{11}$$

$$\frac{\partial E}{\partial r} = 1 - n_e - n - \frac{2\,E}{r}\,, \qquad \textbf{(12)}$$

where, $c = \dfrac{n_{i0}}{n_0\, Z}$, $a = \dfrac{n_{dn}}{w_{pd}}$, $h = \dfrac{Z\, T_i\, D}{T}$, $t_i = \dfrac{T_i}{T_e}$, $t_d = \dfrac{T}{T_e}$ and $l_{di} = \mathrm{v}_{\mathrm{thi}}\, w_{pi}^{-1}$.

Equations (8-12), where dust density is normalised by $\dfrac{n_{i0}}{Z}$, ion density by $n_{i0}$, electron density by $n_{e0}$, time by $w_{pd}^{-1}$ , distance by $\dfrac{l_{di}}{t_i}$ , velocity by $\dfrac{l_{di}\, w_{pd}}{t_i}$, the electric field by $\dfrac{T_e}{e\, l_{di}}$ and the diffusion coefficient by $\dfrac{w_{pd}\, l_{di}^2}{t_i^2}$, copes with a stationary solution as the void necessarily evolves to a stationary state at the final stage.

Equations (9) and (10) may be written respectively in the case of ABH and KIMT forms of the ion drag force as follows,

<u>ABH force</u>

$$\frac{\partial \Psi}{\partial r} = \frac{n}{D\,\mathrm{v}} \left\{ c\, E \left( \frac{a}{b + m^3\, E^3} - 1 \right) + \mathrm{v}\, h\, (-\, n_e - n + 1) - \mathrm{v}\, a \right.$$
$$\left. + \Psi \left[ -\frac{\mathrm{c}}{\mathrm{Z}\, n} + \frac{\mathrm{v}}{n} (\mathrm{v} + h\, E) - \frac{2\, t_i\, D}{l_{di}\, r} \right] + \frac{2\, t_d\, t_i\, \mathrm{v}^2}{\mathrm{Z\, r}} \right\}, \qquad \textbf{(9-b)}$$

$$\frac{\partial \mathrm{v}}{\partial r} = \frac{c\, E}{\mathrm{v}} \left( -1 + \frac{a}{b + m^3\, E^3} \right) - \frac{c\, m\, t_d\, t_i\, \Psi}{n\, \mathrm{v}} - a\,, \qquad \textbf{(10-b)}$$

where, $u = m\, E$, and $a = \dfrac{m\, v_{di}}{m_i\, Z\, v_{in}}$.

KIMT force

$$\frac{\partial \Psi}{\partial r} = \frac{n}{D\,\mathrm{v}} \left\{ c\,E \left( \frac{8\sqrt{2\pi}\, r^2\, n_i\, \mathrm{v}_{thi}\, \Omega}{3\,Z\,\nu_{in}} - 1 \right) + \mathrm{v}\,\eta\,(\,-n_e - n + 1) - \mathrm{v}\,\alpha \right. \\ \left. + \Psi \left[ -\frac{c}{Z\,n} + \frac{\mathrm{v}}{n}\,(\,\mathrm{v} + \eta\,E\,) - \frac{2\,\tau_i\,D}{\lambda_{di}\,r} \right] + \frac{2\,\tau_d\,\tau_i\,\mathrm{v}^2}{Z\,r} \right\}, \qquad \textbf{(9-c)}$$

$$\frac{\partial \mathrm{v}}{\partial r} = \frac{c\,E}{\mathrm{v}} \left( -1 + \frac{8\sqrt{2\pi}\, r^2\, n_i\, \mathrm{v}_{thi}\, \Omega}{3\,Z\,\nu_{in}} \right) - \frac{c\,m\,\tau_d\,\tau_i\,\Psi}{n\,\mathrm{v}} - \alpha\,, \qquad \textbf{(10-c)}$$

where, $\Omega = \left( 1 + \frac{1}{2}\,z\,\tau + \frac{1}{4}\,z^2\,\tau^2\,\Lambda \right)$.

## §.III. Results and discussion

We integrate numerically Eqs. (8-12), the initial conditions, being $\Psi(0) = 0.0001$, $n(0) = 10^{-16}$, $n_e(0) = 0.8$ and $E(0) = 0$, with the following parameters $n_{i0} = 1.86 \times 10^9\ cm^{-3}$; $\nu_{in} = 10^6\ s^{-1}$; $\omega_{pi} = 9 \times 10^5 s^{-1}$ and different grain sizes, viz., $r_d = 120\,nm$ ($\nu_{dn} = 5.97\,10^2 s^{-1}$), $r_d = 350\,nm$ ($\nu_{dn} = 5.07\,10^3 s^{-1}$), $r_d = 700\,nm$ ($\nu_{dn} = 2.03\,10^4 s^{-1}$), $r_d = 1.32\ mm$ ($\nu_{dn} = 7.2\,10^4 s^{-1}$), $r_d = 2.6\ mm$ ($\nu_{dn} = 2.79\,10^5 s^{-1}$). In figures (7-12), different values of the grain radius are considered for two different expressions of the ion drag force, viz, the ABH and the KIMT expressions. In a spherical coordinate system, the three dimensional plot of the dust density may be obtained by way of a rotation around the axis of symmetry, which is not possible with a one-dimensional solution as it is not symmetric around $x = 0$. It appears through the figures that a dust void with a sharp boundary exists provided the grain radius exceeds a critical value, which is different for different expressions of the ion drag force. The dimension of the void decreases with an increasing grain radius. However, for small values of the grain size and when the ABH form of the ion drag force is used, the tendency of the void dimension is towards the increase for increasing grain radius. Moreover, the dimension of the void seems to reach an asymptotic value when the grain radius is increased. The use of different expressions of the ion drag force shows that the ABH force generates a bigger void than the one generated by the KIMT force. For higher grain sizes, no significant difference between the voids generated by the ABH and KIMT forces is seen. In other words, for big grains, the expression derived by Avinash *et al.* leads to acceptable results. Furthermore, in all figures, we notice a density build up at the void boundary (a density peak) which is in agreement with the experiment. Besides, in some situations it is noticed a density corrugation beyond the void, which is in agreement with the experiment also. These are alternate zones of rarefaction and bunching of dust grains.

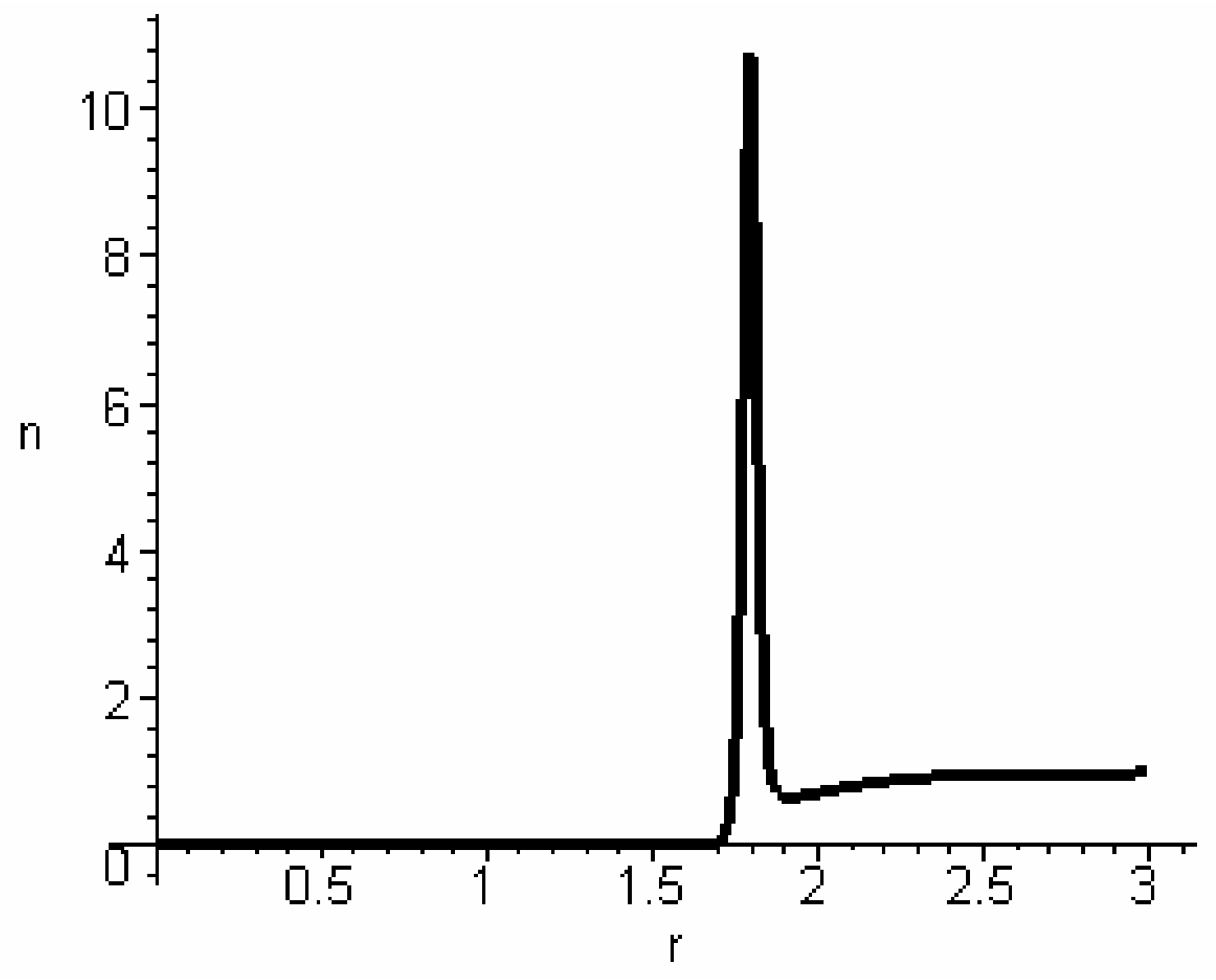

Fig.7. The numerical dust density $n$ versus the position $r$ for $r_d = 350\ nm$

(ABH ion drag force)

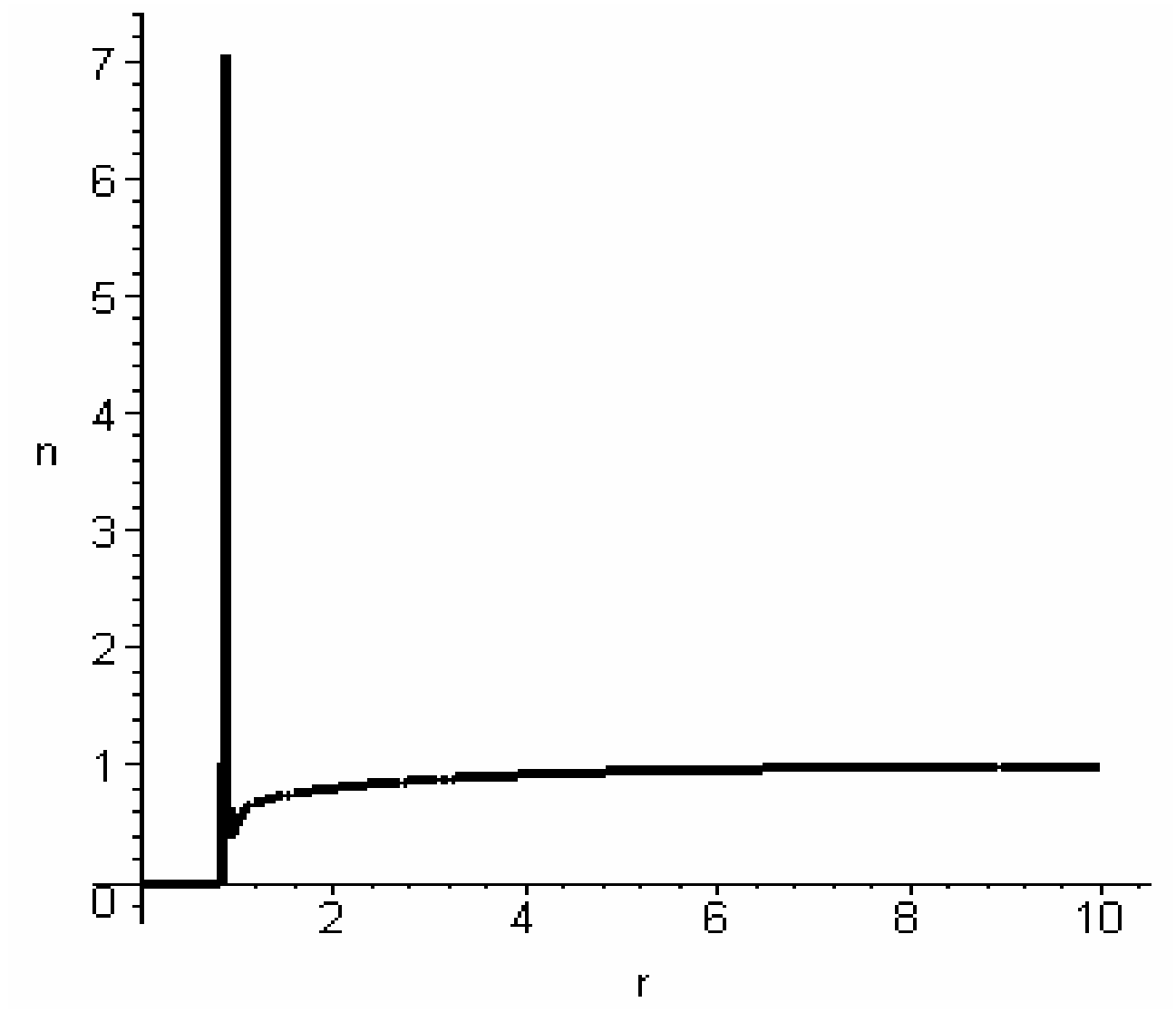

Fig.8. The numerical dust density $n$ versus the position $r$ for $r_d = 700\ nm$

(ABH ion drag force )

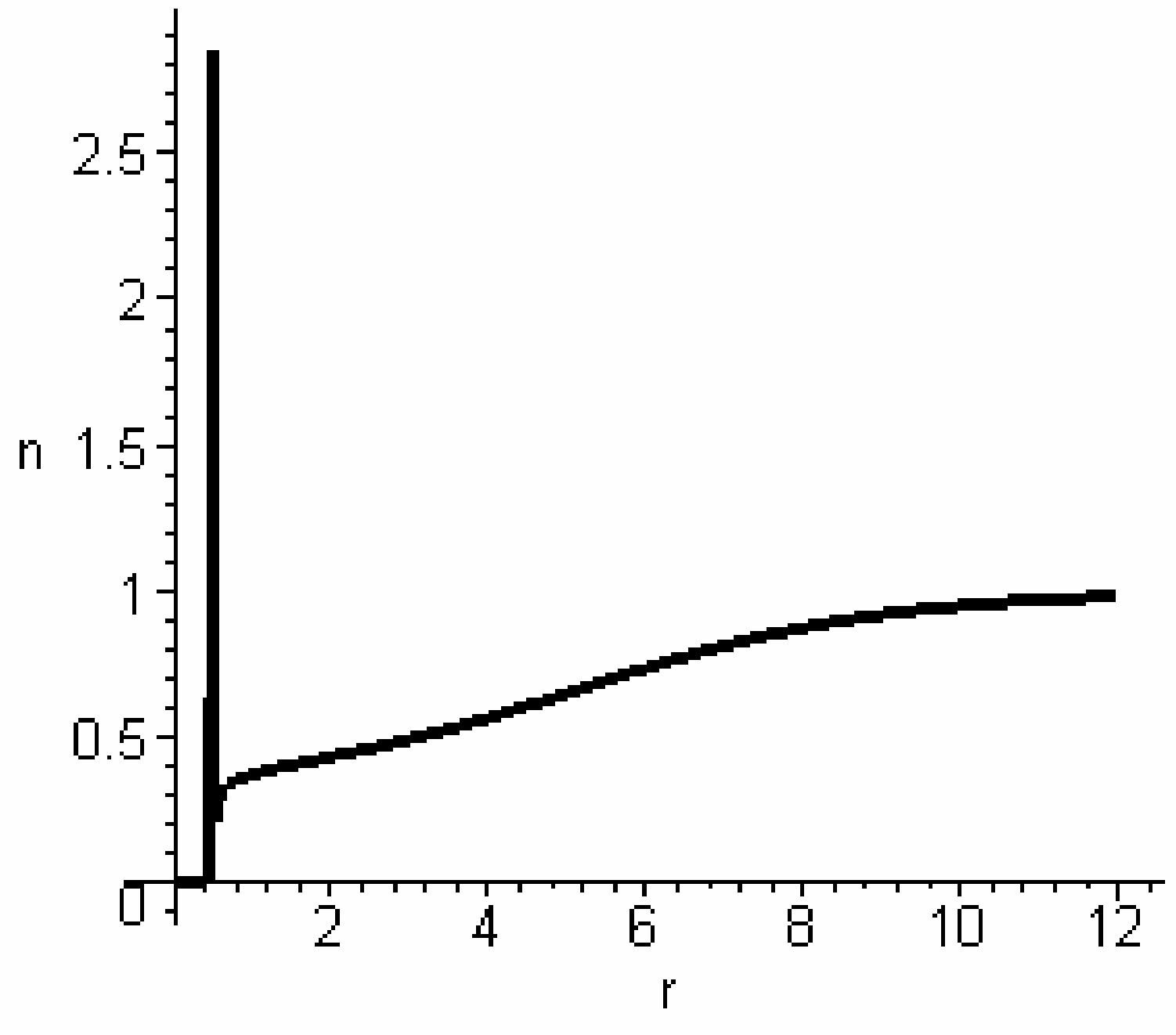

Fig. 9. The numerical dust density $n$ versus the position $r$ for $r_d = 1,320\ nm$

(ABH ion drag force)

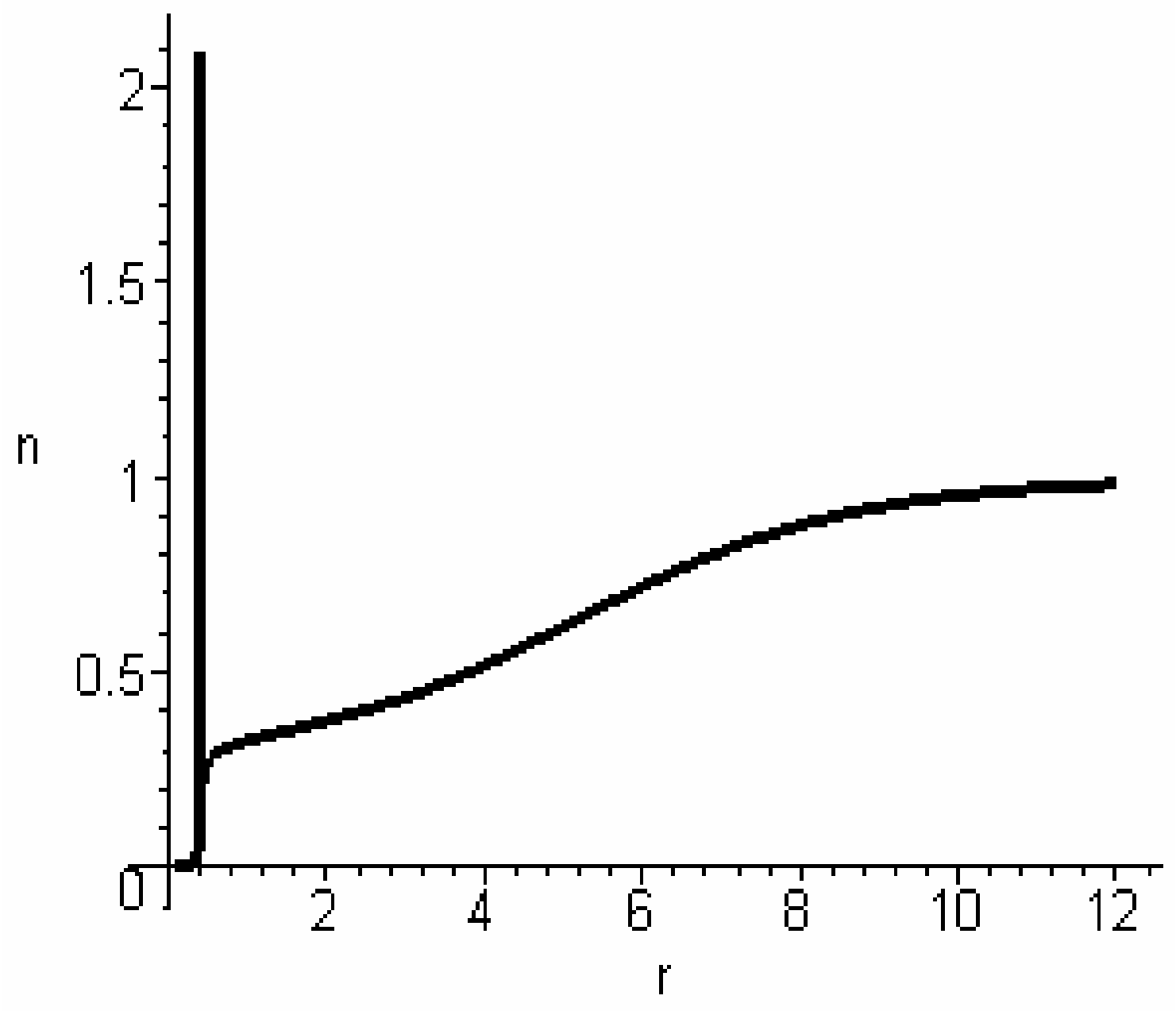

Fig.10. The numerical dust density $n$ versus the position $r$ for $r_d = 2.6\,nm$

(ABH ion drag force)

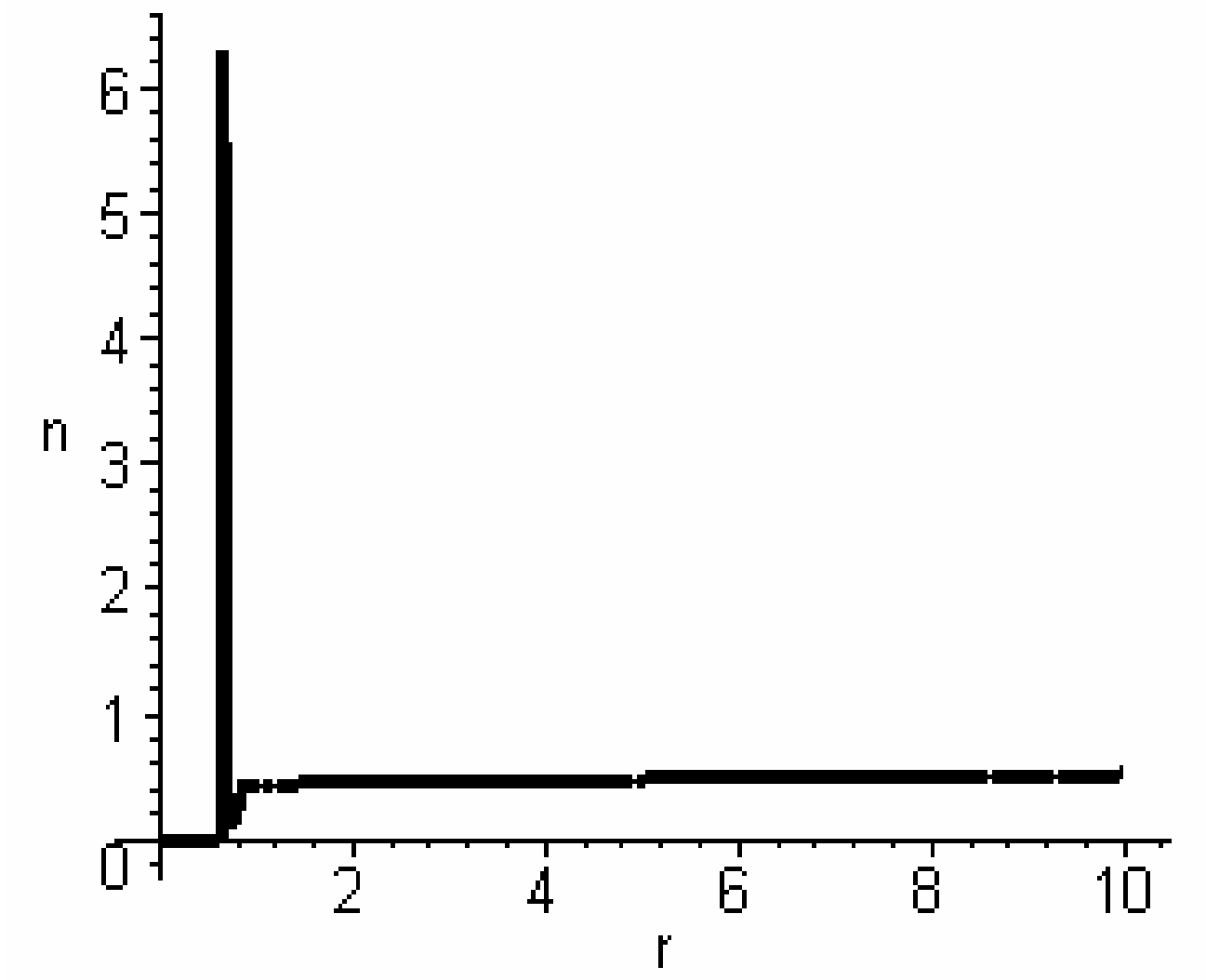

Fig.11. The numerical dust density $n$ versus the position $r$ for $r_d = 700\ nm$

(KIMT ion drag force)

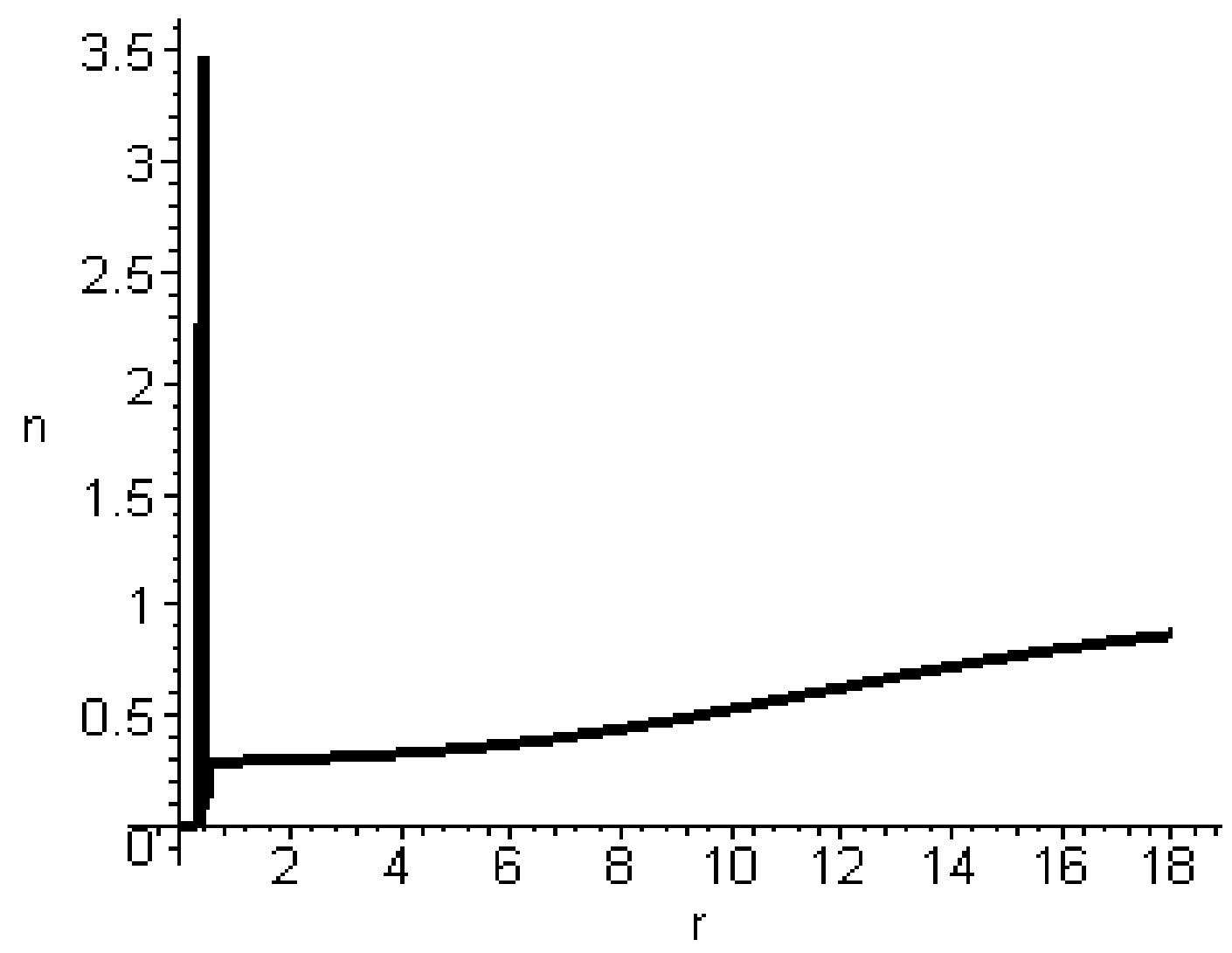

Fig.12. The numerical dust density $n$ versus the position $r$ for $r_d = 1.320\ mm$ (KIMT ion drag force)

## §. IV. Conclusion

The charged dust grains in a complex plasma, exhibit a collective behavior due to their charge, so new modes are excited. Besides, due to their nature the grains rearrange themselves under certain conditions in the plasma. Indeed, when the particulates grow to a certain size, a cloud is formed that undergoes a transition to a turbulent state to lead to a void of dust, i.e., a dust free zone determined by the balance of the electric and the ion drag forces. The void is characterized by a critical size of the grain and a sharp boundary. In order to explain the spontaneous development of the linear instability and the appearance of the void in the final stage, Avinash *et al.* proposed a self-consistent non-linear one-dimensional model that takes into account grain diffusion and collision with the neutrals, and where the non-linear ion drag force has been modeled by an appropriate expression. In this paper, we augmented this model by incorporating the grain drift under the action of the electric field along with an accurate ion drag force derived by Khrapak *et al.* from first principles. The analysis has been conducted in a spherical configuration. The numerical integration of the differential equations revealed the following aspects: a void with sharp boundaries indeed exists, provided the grain radius exceeds a critical value. The dimension of the void is found grain size dependent, though it reaches a stable value for increasing grain size. It has been shown also, that the void generated by the ABH force is bigger than the one excited by the more accurate expression of the ion drag force, viz., KIMT force. For higher values of the grain size, both expressions of the ion drag force lead to voids of comparable dimensions. Furthermore, as shown by experiments, the void boundary is a density peak beyond which the density falls down abruptly and may presents farther, a corrugated profile.

**Acknowledgment**

R.A acknowledges financial support of USTHB (Algeria) and UKZN (South-Africa). This work was initiated at UKZN in 2004.